\documentclass{llncs}

\usepackage{graphicx}
\usepackage[square]{natbib}

\begin{document}

\title{A Framework for Textbook Enhancement and Learning using Crowdsourced Annotations}

\author{Anamika Chhabra\inst{1} \and S. R. S. Iyengar\inst{1} \and Poonam Saini\inst{2} \and Rajesh Shreedhar Bhat \inst{3}}

\institute{Dept of Computer Science and Engineering, IIT Ropar, Punjab, India \\
\email{anamika.chhabra@gmail.com,}
\email{sudarshan@iitrpr.ac.in}
\and
{Dept of Computer Science and Engineering, PEC University of Technology, India }\\
\email{poonamsaini@pec.ac.in}
\and
{Dept of Information Science and Engineering, PESIT, Bangalore, India}\\
\email{rajeshbhatpesit@gmail.com}
}
\maketitle

\begin{abstract}

Despite a significant improvement in the educational aids in terms of effective teaching-learning process, most of the educational content available to the students is less than optimal in the context of being up-to-date, exhaustive and easy-to-understand. There is a need to iteratively improve the educational material based on the feedback collected from the students' learning experience. This can be achieved by observing the students' interactions with the content, and then having the authors modify it based on this feedback. Hence, we aim to facilitate and promote communication between the communities of authors, instructors and students in order to gradually improve the educational material. Such a system will also help in students' learning process by encouraging student-to-student teaching. Underpinning these objectives, we provide the framework of a platform named \textit{Crowdsourced Annotation System} (CAS) where the people from these communities can collaborate and benefit from each other. We use the concept of in-context annotations, through which, the students can add their comments about the given text while learning it. An experiment was conducted on 60 students who try to learn an article of a textbook by annotating it for four days. According to the result of the experiment, most of the students were highly satisfied with the use of CAS. They stated that the system is extremely useful for learning and they would like to use it for learning other concepts in future. 

\end{abstract}

\section{Introduction}\label{sec:Introduction}

The emergence of Web 2.0 and Web 3.0 technologies has facilitated development of various online environments equipped with the features of social learning\citep{richardson2010blogs}\citep{Kesim2007}\citep{Kolbitsch2006}. Due to the ubiquity of the internet, various resources such as electronic textbooks, online forums, MOOCs etc. are accessible at one’s fingertips for interactive and easy learning \cite{martin2012will}. Also, the students have become more comfortable reading material online.  Although the online teaching/learning method has successfully staggered the enrolment of students into such courses and portals, the level of participation on the part of students is not satisfactory\cite{Anderson}. The possible reasons may include inadequate preparation of online educational systems in terms of quality of content as well as the interface\cite{Vonderwell2005}. Unfortunately, the education system has no provision of making use of the feedback from interaction of students during the learning process. This leads to the creation of textbooks which are not in sync with the actual requirements of students for active learning\cite{Fung2011}. There is an immense requirement to eliminate the communication gap between the students, instructors and the authors by providing them a common platform, where the people from these diverse communities can benefit from each other. \\
We aim to provide the framework of an environment, which the authors can use to reconsider parts of their textbook modules needing improvement or further explanation, after observing the students’ interactions when they try to understand some part of their textbook. This type of environment will not only help the authors in getting constructive feedback about their own content, but will also help the instructors to provide a virtual classroom type of environment with the features of discussions, brainstorming and feedback to the students.  Also, they can closely observe the interactions taking place among the students and can intervene whenever required. It will help the students in learning various concepts with the help of other students which will enable student-to-student teaching. Hence, such a portal will be beneficial for three types of stakeholders: \textit{Authors, Students} and \textit{Instructors}.  \\
The use of discussion forums is the most prevalent method of collecting knowledge from the people in a crowd. Also, people are known to learn better when they participate in discussion forums based websites\cite{Davies2005}\cite{Deslauriers2011a}. Such websites also improve the engagement level of the participants\cite{Palmer2008}. However, there is one constraint which is associated with the portals based on discussion forums. Due to the non-linear structure of such forums, the user sometimes gets deviated from whatever information he was originally searching for\cite{Thomas2002}. An in-place annotation tool can outperform traditional forums as a medium for discussion of classroom material. It helps the students to remain in the flow instead of losing context on a different forum\cite{Cadiz2000}\cite{Zyto2012}. Hence, we suggest the use of annotations in the document margins, where the collaboration will happen. \\
Keeping in view all these objectives, we propose the framework of a portal named Crowdsourced Annotation System (CAS), which provides a facility to the authors to add textbook modules to the content repository, a facility to the instructors to choose a module from the repository and authorize a group of students, and a facility of annotating any part of the text to the students. Chhabra et al.\cite{chhabra2015presence} verify the presence of variance of expertise on portals such as CAS, which creates an ecosystem among the categories of annotators. The authors demonstrate that it is this ecosystem, that acts as an impetus in the knowledge building process. \\
In order to verify the efficacy of CAS, we conducted an experiment with a group of students who annotated a textbook module while learning it through the portal. Later, a feedback analysis for CAS was also done, in which, most of the students agreed that the portal helped them a great deal in understanding the given textbook article.
\section{Related Work}\label{sec:related}
There has been a significant amount of interest to bring in ICT technology in education and learning in general. The contributions in this line are increasingly getting interdisciplinary, with academic communities ranging from educational psychologists\cite{Cress2008} to computer scientists\cite{Zyto2012} getting involved, thus making this domain of research transdisciplinary.
\subsubsection{Education and crowdsourcing}
Corneli and Mikroyannidis\cite{Chapter2012} investigated the concept of crowdsourcing in education through an analysis of case studies on two open online learning communities- Peer 2 Peer University and PlanetMath.org. The authors analyzed various roles played by the individuals involved and also envisioned future roles that could be created at these organizations. They found that crowdsourcing model provides all the roles which are found in the traditional environments, as well as offers some additional richness due to the social networking tools. Weld et al.\cite{Weld2012a} investigated the potential of crowdsourcing methods for improving online personalized education. The authors discussed various challenge areas that offer promise of rapid progress and substantial impact. Porcello and Hsi\cite{Porcello2013} discussed how multiple user communities and online platforms might be coordinated to provide effective experiences in online learning environments.
\subsubsection{Annotation Systems}
In order to eliminate the problems that discussion forums pose, a few systems have been built by various researchers including \textit{CAMILE}\cite{Guzdial2000}, \textit{WebAnn}\cite{Marshall2002}, \textit{NB System}\cite{Zyto2012} and \textit{PAMS 2.0}\cite{Su2010} etc. CAMILE is based on a specific type of computer-mediated anchored discussion forum, where the threads are linked to each other, but it still requires the users to go to another context. Although WebAnn provides the advantages of in-context annotations, yet it was used just as a record-keeping tool due to discomfort of users in reading online material. NB System helped in alleviating the problems faced by WebAnn. It is an online portal where users can read and annotate any part of a pdf document. PAMS 2.0 is another annotation system based on Web 2.0 technology. In their work, the authors demonstrated that the effect of annotation on learning achievements becomes more substantial with the use of sharing technique. 
\subsubsection{Textbook Creation}
In order to address the problem of high cost of text-books for Higher Education in Latin America, Ochôa, Sprock and Silveira\cite{Ochoa2011} worked on the creation of collaborative open text-books that could be freely and legally copied, printed, modified and distributed to students. Baraniuk et al\cite{Baraniuk2002} developed a portal named \textit{Connexions} for customized textbook creation. However, our work is significantly different from theirs in the sense that they provide various textbooks in the form of modules. The students and instructors can develop customized textbooks as per their requirements by combining these modules. However, we make use of the annotations, which the students add to the given part of the textbook in order to understand it. The kind of interaction taking place among the students gives an idea to the authors about the parts to be reconsidered. Also, this kind of platform helps in the learning process of the students.
\section{Crowdsourced Annotation System(CAS)-A Proposal}\label{sec:CAS}
\begin{figure}
\centering
\includegraphics{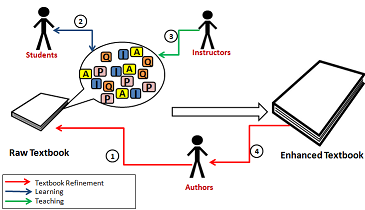}
\caption{Schematic View of the Textbook Refinement Process. Red, Blue and Green lines indicate the purpose for which authors, students and instructors respectively will make use of the system.}
\label{fig:Scheme}
\end{figure}

\begin{figure}
\centering
\includegraphics{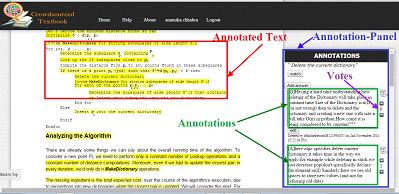}
\caption{A snapshot of CAS. (The Crowdsourced Annotation System (CAS) can be accessed at http://115.248.248.12/CrowdSourced/)}
\label{fig:CAS}
\end{figure}

We propose a technique of textbook creation and refinement coupled with the advantages of annotation systems. By providing an integrated platform in which existing content is combined with an interactive environment where the people in the crowd collaborate, the quality of material can be immensely improved on a gradual ongoing basis. Moreover, using the power of crowd into the content generation process will improve the execution speed along with the generation of diverse ideas, thereby, improving the quality of content. The motive is to allow authors to reconsider the portions of their text to be modified based on the feedback from the crowd. Therefore, in a way, everybody in the crowd becomes an author of the text in consideration. Also, it will successfully utilize the intelligence of the great masses of the people, which will be a mixture of people possessing various types of expertise.\\
Essentially, the proposed framework has three major interdependent components: (a) a repository of content added by the authors, (b) an interface to add in-context annotations to the content, and (c) a mechanism for the collaboration of annotators. Essentially, the system is an integration of three major notions: crowdsourcing, online textbook assessment and collaborative learning. All the stakeholders of the system collaborate on the creation of knowledge in an extremely dynamic environment. Figure~\ref{fig:Scheme} shows the raw textbook posted by the author being annotated by the students. The instructors closely observe the annotations posted, and intervene to help the students, whenever required. The textbook after having been annotated, provides enormous amount of information to the author so as to make further modifications to the text as per the students expectations, and can easily come out with a modified version.\\
We created a web-based annotation system named as Crowdsourced Annotation System (CAS) which depicts the proposed framework. In CAS, a group of students are authorised to access a particular article of a particular textbook by the instructor. The students communicate with each other by adding annotations. The annotations added, are attached to the corresponding text of the article. The text for which there are annotations, is displayed highlighted and clicking on this highlighted part displays all the annotations for this text. Also, the annotations are displayed on the same page in the document margins, which every other student can view and comment. The students can add four types of annotations, namely, \textit{Questions} (A), \textit{Answers} (A), \textit{Insights} (I) and \textit{Pointers} (P). Question and Answer type of annotations provide the advantages of a discussion forum, however, alleviating the problems of a forum by displaying the discussion in the same context. The annotators can add any extra information about some text through the use of Insight annotations. Pointer type of annotations provide a facility to refer to some outside relevant material regarding the text. These four type of annotations help the students in properly communicating their comments about the article being studied. A snapshot of the CAS is shown in Figure~\ref{fig:CAS}.
\section{The Experiment}\label{sec:expt}
In order to understand how students' annotations emerge over time, and how these can be used in the textbook refinement process, we conducted an experiment\footnote{In\cite{chhabra2015presence}, Chhabra et al. verify the existence of an ecosystem among the annotators' categories using this experimental data.} on a group of 60 CS students from the \textit{Indian Institute of Technology Ropar, India}. The students were authorized to annotate an article on ``Randomized Closest Pair Problem'' from a reference book\cite{kleinberg2006algorithm} for their course on Algorithm Design. This article was new to them and was not easy for them to understand by independent study. The aim for the students was to understand this article with the help of each other. They annotated the article for a duration of four days. All the transactions 
happening among the students were stored in a database for analysis. 
Table~\ref{fig:Stats} shows various items of statistics regarding the experiment, including the number of annotations, percentage distribution, maximum, minimum, mean and standard deviation for the four categories of annotations.
\begin{table}
\centering  
\includegraphics{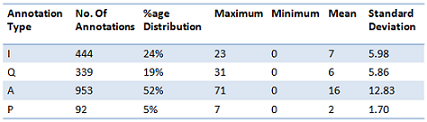}
\caption{Number, Distribution, Max, Min, Mean and SD of Annotation Types (Our analysis data has been kept open at the CAS portal.)}
\label{fig:Stats}
\end{table}
\\Also, we recorded certain other parameters of the students' interaction:
\begin{itemize}
\item There were a total of 15932 reading entries i.e. number of annotations read by the students.  
\item There was also a feature of up voting or down voting of the annotations. There were a total of 811 voting entries.
\item A student could click on a watch button attached to every annotation that he added, so as to be informed about the kind of responses that he got in future for his own annotations. There were a total of 66 watcher entries.
\item A session denotes a collection of interactions taking place within a particular time period. There were 875 sessions with an average duration of 15.49.
\item There were a total of 3159 pageviews which represents the total number of pages accessed during the period of 4 days.
\end{itemize}
\section{User Assessment of CAS}\label{sec:assessment}
\begin{figure}
\centering
\includegraphics{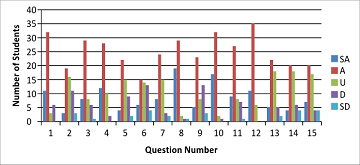}
\caption{Distribution of Responses}
\label{fig:Assessment}
\end{figure}

Emerging information technology cannot be effective if it is not accepted and used by potential users. Therefore, at the end of the session, the students were asked to provide their feedback about CAS through a web-based form. The form used \textit{Technology Acceptance Model (TAM)} to explore the \textit{`Perceived Usefulness (PU)'} and \textit{`Perceived Ease of Use(PEOU)'} of CAS\cite{Fred1989}. TAM is one of the most successful models to assess the user acceptance of emerging information technology among practitioners and academicians. The model attempts not only for prediction but also for the explanation to help researchers and practitioners identify why a particular system may be unacceptable and pursue appropriate steps. PU tells prospective user’s subjective probability that using a specific application system will increase his or her job performance within an organizational context. PEOU tells the degree to which the prospective user expects the target system to be free of effort.\\
The universal efficacy of CAS was also established in the feedback from the participants. They reported that the introduction of various types of annotations (i.e. I, Q, A, P) in CAS was useful in sharing the knowledge in the group. Displaying the annotations on the same page as the text was useful in bringing the students’ attention towards annotations added by others, which further enhanced their knowledge.
Apart from these two dimensions, i.e., PU and PEOU, the questions were asked on one more dimension named \textit{`Satisfaction (SAT)'}, to measure the degree of contentment among the participants. The Feedback Form included fifteen questions using a five-point Likert scale (strongly agree; agree; undecided; disagree; strongly disagree). The questions were classified as: (1) PU (7 Questions); (2) PEOU (5 Questions); (3) SAT (3 Questions). Out of 64 participants, 52 had filled the form.  On a scale assigning the values as: 5 for ‘Strongly agree’, 4 for ‘Agree’, 3 for ‘Undecided’, 2 for ‘Disagree’ and 1 for ‘Strongly Disagree’, the participants reported on a mean in the range of 3.12 and 4.25 for all the fifteen questions.\\
Following are the 15 questions asked in the assessment:
\begin{enumerate}
\item I think the introduction of various types of annotations (i.e. I, Q, A, P) in CAS was useful in sharing the knowledge in the group. 
\item I think with the use of CAS, I was able to understand the given text in less time.
\item I think displaying the annotations on the same page as the text was useful in bringing attention towards annotations added by others.
\item I think CAS was helpful in exchanging my thoughts across others, while reading.
\item I  think I have been able to understand the text in a better way through the use of CAS, as compared to what I would have done, without the use of such an interface.
\item I think that the voting of annotations was useful in scrutinizing high quality annotations out of all the annotations.
\item I think such an annotation system is more useful than a simple discussion forum.
\item I think it was easy to add a new annotation to the text.
\item I think it was easy to view the annotations added by others in the right panel.
\item I think it was easy for me to understand how to use CAS.
\item I think the Graphical Interface was quite easy to work with
\item I think due to the introduction of various categories of annotations, it was easy for me to put across my point on the interface.
\item I am thoroughly satisfied with use of CAS for reading an article.
\item I would like to use this interface for reading other texts also.
\item I would like to use this interface in future.
\end{enumerate}
Figure~\ref{fig:Assessment} shows the distribution of various responses of the 15 questions by all the 52 students. 
On an average, 82.69 percent students gave `Strongly Agree' or `Agree' as their answers.\\
We also analysed the suggestions/comments that the students were asked to fill, along with the ratings. Following are some of the comments:
\textit{
\begin{itemize}
\item The idea of having such a system is great, and CAS was good.
\item CAS is very good source to understand texts in a better way.
\item It was user friendly and very useful for understanding while reading the article.
\item I would like to test this on further topics.
\item Overall it was a good learning platform where you can easily share your understandings with other people in the group, ask questions when required and get the answers and point of view of other people. It is a good tool to assess and evaluate your understanding of the concept.
\item CAS is very good source to understand texts in a better way.
\item The idea of having such a system is great, and CAS was good.
\item Through this, learning was very easy and quick and also enjoyable.
\item Everything is really good..be it the user-interface or the whole concept of IT... Infact..(sic) I with some of my intellectual friends seriously want to study through this medium where exchange of ideas is so easy.
\item The CAS system was good enough to learn /analyse the text in less time.
\item The site was perfectly built and was very useful to fully understand the topic.
\end{itemize}
}
\section{Conclusion}\label{sec:conclusion}
In this work, we presented the framework of an integrated platform, CAS, which can be used by the authors to collect feedback about their textbooks. The textbooks which are created for the students, can be best judged by the students only when they try to read them and provide their comments. Also, it provides the students a virtual learning environment to help each other understand the content. For this purpose, we developed and executed an experiment involving text annotations by students on CAS. An online feedback of CAS was later conducted to evaluate how students come to accept and use the system. As per the feedback, most of the students mentioned that they would like to use the system for learning in future. \\
We believe that a system like CAS will help a great deal in improving the content of the textbooks by getting feedback from the interaction of the students. It will also influence the creation of textbooks in future by showing an integration of text, collaboration and opinion. The usefulness of the interface will be even beyond the knowledge refinement. It will be fit for use in many other domains as well, where any type of collaboration or decision making based on the existing facts is required. For example, the system will be usable for the review of research papers created by various researchers, for collaboration of doctors on a complex medical case etc. The platform will also help as a perfect infrastructure for the reviewers of any conference/journal to discuss their views about a particular paper.

\bibliographystyle{splncs}
\bibliography{lncs_iccci}

\begin{thebibliography}{10}

\bibitem{richardson2010blogs}
Richardson, W.:
\newblock Blogs, wikis, podcasts, and other powerful web tools for classrooms.
\newblock Corwin Press (2010)

\bibitem{Kesim2007}
Kesim, E., Agaoglu, E.:
\newblock {A Paradigm Shift in Distance Education: Web 2.0 and Social
  Software.}
\newblock Online Submission (2007)  66--75

\bibitem{Kolbitsch2006}
Kolbitsch, J., Maurer, H.:
\newblock {The Transformation of the Web : How Emerging}.
\newblock Journal of Universal computer Science \textbf{12} (2006)  187--213

\bibitem{martin2012will}
Martin, F.G.:
\newblock Will massive open online courses change how we teach?
\newblock Communications of the ACM \textbf{55} (2012)  26--28

\bibitem{Anderson}
Anderson, A., Huttenlocher, D., Kleinberg, J.:
\newblock {Engaging with Massive Online Courses}.
\newblock In: WWW'14 ACM. (2014)

\bibitem{Vonderwell2005}
Vonderwell, S., Zachariah, S.:
\newblock {Factors that influence participation in online learning}.
\newblock Journal of Research on Technology in Education \textbf{5191} (2005)
  213--230

\bibitem{Fung2011}
Yan, Y., Fung, G.:
\newblock {Active learning from crowds}.
\newblock Proceedings of the 28th International Conference on Machine Learning,
  (2011)

\bibitem{Davies2005}
Davies, J., Graff, M.:
\newblock {Performance in e-learning: online participation and student grades}.
\newblock British Journal of Educational Technology \textbf{36} (2005)

\bibitem{Deslauriers2011a}
Deslauriers, L., Schelew, E., Wieman, C.:
\newblock {Improved learning in a large-enrollment physics class.}
\newblock Science (New York, N.Y.) \textbf{332} (2011)  862--4

\bibitem{Palmer2008}
Palmer, S., Holt, D., Bray, S.:
\newblock {Does the discussion help? The impact of a formally assessed online
  discussion on final student results}.
\newblock British Journal of Educational Technology \textbf{39} (2008)
  847--858

\bibitem{Thomas2002}
Thomas, M.:
\newblock {Learning within incoherent structures: the space of online
  discussion forums}.
\newblock Journal of Computer Assisted Learning \textbf{18} (2002)  351--366

\bibitem{Cadiz2000}
Cadiz, J.J., Gupta, A., Grudin, J.:
\newblock {Using Web Annotations for Asynchronous Collaboration Around
  Documents Using Web Annotations for Asynchronous Collaboration Around
  Documents}.
\newblock In Proceedings of the 2000 ACM conference on Computer supported
  cooperative work (2000)  309--318

\bibitem{Zyto2012}
Zyto, S., Karger, D., Ackerman, M., Mahajan, S.:
\newblock {Successful classroom deployment of a social document annotation
  system}.
\newblock Proceedings of the SIGCHI Conference on Human Factors in Computing
  Systems, ACM (2012)  1883--1892

\bibitem{chhabra2015presence}
Chhabra, A., Iyengar, S., Saini, P., Bhat, R.S.:
\newblock Presence of an ecosystem: An answer to" why is whole greater than the
  sum of its parts" in the knowledge building process.
\newblock arXiv preprint arXiv:1502.06719 (2015)

\bibitem{Cress2008}
Cress, U., Kimmerle, J.:
\newblock {A systemic and cognitive view on collaborative knowledge building
  with wikis}.
\newblock International Journal of Computer-Supported Collaborative Learning
  \textbf{3} (2008)  105--122

\bibitem{Chapter2012}
Corneli, J., Mikroyannidis, A.:
\newblock {Crowdsourcing Education on the Web: A Role based Analysis of Online
  Learning Communities}.
\newblock IGI Global (2012)

\bibitem{Weld2012a}
Weld, D.S., Adar, E., Chilton, L., Hoffmann, R., Horvitz, E., Koch, M., Landay,
  J., Lin, C.H.:
\newblock {Personalized Online Education --- A Crowdsourcing Challenge}.
\newblock (2012)

\bibitem{Porcello2013}
Porcello, D., Hsi, S.:
\newblock {Crowdsourcing and curating online education resources}.
\newblock Science Education (2013)

\bibitem{Guzdial2000}
Guzdial, M., Turns, J.:
\newblock {Effective discussion through a computer-mediated anchored forum}.
\newblock The journal of the learning sciences \textbf{9} (2000)  437--469

\bibitem{Marshall2002}
Marshall, C.C., Brush, A.J.B.:
\newblock {From personal to shared annotations}.
\newblock CHI '02 extended abstracts on Human factors in computing systems
  (2002)  812--813

\bibitem{Su2010}
Su, A.Y., Yang, S.J., Hwang, W.Y., Zhang, J.:
\newblock {A Web 2.0-based collaborative annotation system for enhancing
  knowledge sharing in collaborative learning environments}.
\newblock Computers \& Education \textbf{55} (2010)  752--766

\bibitem{Ochoa2011}
Och\^{o}a, X., Sprock, A., Silveira, I.:
\newblock {Collaborative Open Textbooks for Latin America–the LATIn Project}.
\newblock Technology (2011)

\bibitem{Baraniuk2002}
Baraniuk, R., Burrus, C.:
\newblock {Connexions: DSP education for a networked world}.
\newblock IEEE (2002)

\bibitem{kleinberg2006algorithm}
Kleinberg, J., Tardos, {\'E}.:
\newblock Algorithm design.
\newblock Pearson Education India (2006)

\bibitem{Fred1989}
Davis, F.D.:
\newblock {Perceived Usefulness , Perceived Ease Of Use , And User Accep}.
\newblock MIS quarterly (1989)  319--340

\end{thebibliography}
\end{document}